\documentclass[11pt]{article}

\usepackage[normalem]{ulem}
\usepackage{amsmath, amssymb,amsthm,bm,bbm,fullpage}
\usepackage{mathrsfs}
\usepackage{cite}
\usepackage{colortbl}
\usepackage[all]{xy}
\usepackage{epsfig}
\usepackage{subfigure}
\usepackage{graphics}
\usepackage{multirow}
\usepackage{lineno}
\usepackage{setspace}
\usepackage{caption}

\usepackage{graphicx}

\usepackage{lipsum}
\usepackage{amsfonts}
\usepackage{graphicx}
\usepackage{epstopdf}
\usepackage{algorithmic}
\usepackage{hyperref}
\ifpdf
  \DeclareGraphicsExtensions{.eps,.pdf,.png,.jpg}
\else
  \DeclareGraphicsExtensions{.eps}
\fi

\theoremstyle{remark} 
	
	\doublespace

\begin{document}

\begin{centering}
{\huge
\textbf{Incorporating Computational Challenges into a Multidisciplinary Course on Stochastic Processes}
}
\bigskip
\\
Mark Jayson Cortez$^{1}$, Alan Eric Akil$^{1}$, Kre\v{s}imir Josi\'{c}$^{*,1,2}$, Alexander J. Stewart$^{*,3}$
\\
\bigskip
\end{centering}
\begin{flushleft}
{\footnotesize
$^1$ Department of Mathematics, University of Houston
\\
$^2$ Department of Biology and Biochemistry, University of Houston
\\
$^3$ School of Mathematics and Statistics, University of St Andrews, St Andrews, UK
\\
$^*$ E-mail: kresimir.josic@gmail.com, ajs50@st-andrews.ac.uk
}
\end{flushleft}

\noindent\textbf{ Quantitative methods and mathematical modeling are playing an increasingly important
 role across disciplines. As a result, interdisciplinary mathematics courses are  
 increasing in popularity.  However, teaching such courses 
 at an advanced level can be challenging. Students often arrive with different mathematical 
 backgrounds, different interests, and divergent reasons for wanting to learn the material.  Here we describe
 a course on stochastic processes in biology, delivered between September  and December 2020 to a mixed audience of mathematicians and biologists.
 In addition to traditional lectures and homeworks, we incorporated a series of weekly computational challenges into the course. These challenges served to familiarize students with the main modeling concepts, and provide them with an introduction on how to implement them in a research-like setting. In order to account for the different academic backgrounds of the students, they worked on the challenges in
 small groups, and presented their results and code in a dedicated discussion class each week. We discuss our experience designing and implementing an element of problem-based learning in an applied mathematics course through computational challenges. We also discuss
 feedback from students, and describe the content of the challenges presented in the course. 
 We provide all materials, along with example code for a number of challenges.   
}
\\

\section{Introduction}

Quantitative methods are of growing importance across many fields, as access to big data and powerful computational tools become the norm. And so familiarity with mathematical modeling is increasingly valuable for researchers from all backgrounds. 
Mathematical modeling requires a combination of skills, to successfully bring together discipline specific knowledge of the processes being described, and the mathematical knowledge required to properly formulate a model. Researchers need access to mathematical tools to determine if a model is well-posed, to analyze
its behavior, and to determine solutions when possible. Just as important, interpretation and analysis of a model often requires familiarity with computational methods to implement it, and statistical methods to understand and analyze
its output.   

A course on mathematical modeling thus requires introducing students to a variety of 
mathematical ideas, and domain-specific concepts. This becomes particularly challenging when
teaching an advanced undergraduate or graduate modeling course aimed at an interdisciplinary audience.
Given the range of tools and ideas required to formulate, implement and analyze all 
but the simplest of toy models, such courses need to be organized differently from more
classical applied mathematics courses.   

Here we describe an approach to delivering such a course, implemented between September 2020 and December 2020 at the University of Houston. We developed an introductory graduate course on 
stochastic processes in biology, for an interdisciplinary audience of biologists and applied mathematicians, incorporating an element of problem-based learning \cite{Hmelo04,Hung08,Savery15,Wood03} 
in the form of computational challenges. 
We designed a series of nine sets of such 
challenges, which were assigned to groups of students to tackle in Python. These
challenges were integrated into the course, alongside more traditional 
lectures and homeworks. The challenges in each set were thematically related to each other, and
to the lectures that immediately preceded them. Each challenge required implementation of a mathematical model making use of techniques taught in class, but required students to explore the models in a more open-ended way than in homeworks. Students tackled the challenges
in groups, and presented their code and results in a dedicated class following submission. The ensuing discussions
allowed us to revisit aspects of the material covered in lectures, talk about concrete implementation 
questions and bug fixing, and connect the material
to current questions and methods in biology, and to the practical challenges of implementing mathematical models in a research setting. Translating 
algorithms discussed in class into models of concrete biological processes also offered students a view
of the role of stochastic processes in biology, allowing them to develop intuitions for the biological relevance of concepts such as a stationary distribution, or a first passage time. 

The more open-ended nature of these programming challenges allowed us to go beyond what students typically learn in an
introductory course. Indeed, as the complexity of the challenges increased, it was impossible
to fully specify the modeling approach. This allowed us to discuss how some results depend on 
the modeling choices made by the student or researcher. 
An important consequence of such open-endedness is that the computational challenges did not come with a simple set of right answers. While this was uncomfortable for some students, it also reflects a difficulty faced by many students as they transition from undergraduate courses to independent research and problem solving in graduate school or outside of academia. In contrast to a typical applied mathematics course, the computational challenges helped introduce students to simple heuristics for model validation when no analytical solution is available, and encouraged students to think more carefully about whether computational results made sense given the modeling assumptions, and whether the assumptions made sense given the observed results.

Students tackled the computational challenges in small groups, which were assigned at the start of each challenge. These groups necessarily contained individuals with different levels of programming skill, and mathematical and biological knowledge. We discuss the efficacy of using this kind of small group learning~\cite{Lou01,springer99} 
as part of an applied mathematics course, in particular the challenges of maintaining equitable division of labor, and of maintaining successful group interactions under conditions of remote learning imposed by the COVID-19 pandemic. 

In a repository accompanying this article. provide a total of nine challenges as supplementary material to the paper, which can be used in other courses as written, or used as a basis for the development of similar challenges in other disciplines. Each challenge is broken into 4 interrelated 
group assignments, and  we provide complete sample solutions to three challenges, including 
Python code. We also provide a summary of feedback from the students and discuss planned points of improvement when the course is given in the future.

Our approach of using small group learning and computational challenges to bridge the gaps between an interdisciplinary audience, and to introduce students to some of the practical realities of applied mathematics research, can be replicated by others, and can be easily adapted to fit alternate sets of research interests.


\section{Course Overview}
\label{sec:main}
Biological processes are inherently stochastic~\cite{faisal08,wilkinson09}. While deterministic models 
can offer valuable insights into the function and behavior of living systems, 
they do not capture the effects of randomness and variability that characterize and often drive many biological processes. 
Hence, a familiarity with stochastic processes is vital for anyone seeking to build models of biological systems and simulate their behavior numerically. The goal of our course is to teach students how to use the tools of the mathematical theory of probability and 
stochastic processes to develop, analyze, and implement  models of living systems.  

Our course met for 80 minutes twice a 
week for 14 weeks, but our approach could be easily adjusted to fit different schedules. Due to the COVID-19 pandemic, our class took place remotely via Zoom, and so the need to maintain student engagement was an important issue. To address this,
we broke the class into two 25 minute periods of instruction, separated by 30 minutes devoted to 
student presentations and discussions of computational challenges.

We chose the mathematical topics for the course so as to be of use to a wide audience
of graduate students in applied mathematics, biology, biomedical engineering and related disciplines. Our audience in fall 2020 
consisted of 
16 PhD students (11 in Applied Mathematics, 4 in Biology, and 1 in  Biomedical Engineering).
The prerequisites included  differential equations, linear algebra, as well as undergraduate level probability. We also assumed
that students have some experience programming in MATLAB or Python.  

We asked students to use a Git repository to facilitate
sharing of code related to computational challenges. Two months before the course began we 
checked with students individually to see if they have the requisite background. 
We pointed students with a weaker background in programming to introductory online Python courses \cite{python1}, 
and asked them to complete them before the start of our course. In particular, we 
recommended a tutorial on agent based modeling, as it introduced a number of ideas used in the course \cite{agent_based}. 
We set up a Slack team to communicate with students before and during the course, to facilitate group discussion and keep a clear record of problems that arose and their solutions.  

\begin{figure}[!htpb]
    \centering
    \includegraphics[width=1.0\linewidth]{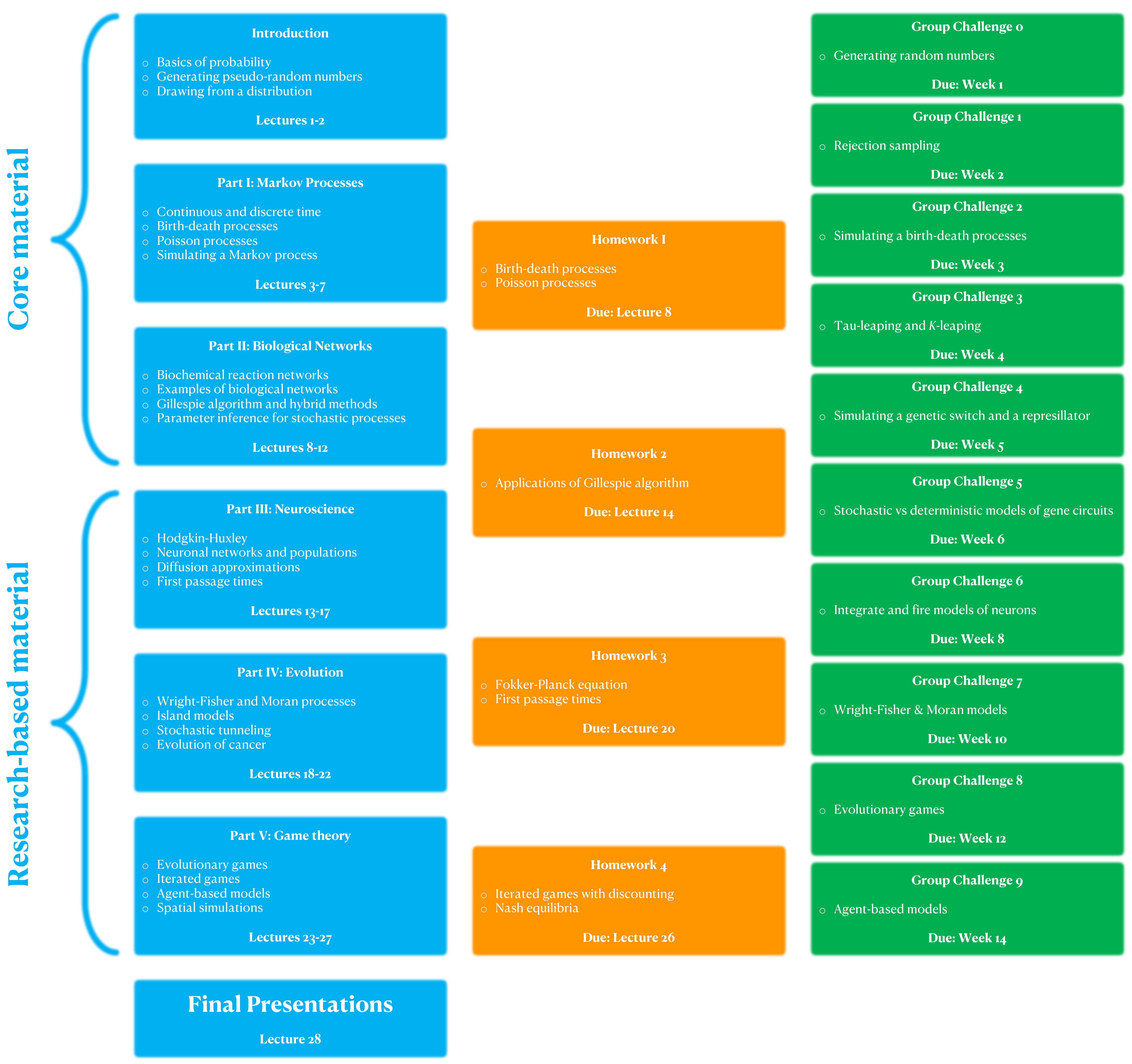}
    \caption{Summary of the course structure. Lectures (blue) were divided into five sections plus an introduction. In each section, key models and concepts related to the topic were introduced, alongside the tools required to analyse and simulate them. Each homework (orange) covers six lectures worth of material. Group computational challenges (green) occur either every week (challenge 0-5) or every two weeks (challenge 7-9) as the challenges become more advanced, and thus cover either two or four lectures worth of material.}
    \label{course_structure}
\end{figure}

An overview of the course is provided in Fig.~\ref{course_structure}, and a detailed syllabus with a complete list of topics covered in the lectures and homeworks
 can be found in the GitHub repository accompanying this article. 
 
The course contained two types of material (Fig.~\ref{course_structure}) with core material mostly delivered first, covering a review of probability, followed by a discussion of Markov processes with discrete and continuous space variables, diffusion processes, stochastic differential equations, Wiener and Ornstein-Uhlenbeck processes, and point processes. 
The discussion of these mathematical concepts was kept
at a practical level, driven by examples and the ultimate goal of simulating 
stochastic models of living systems. We were also inspired by D. Gillespie's dictum 
that ``one’s knowledge of any dynamical system is deficient unless one knows a valid way to numerically simulate that system''~\cite{gillespie91}.  
Thus, as far as possible, we used 
an algorithmic approach in introducing stochastic processes motivated by N. Higham's articles in this journal~\cite{Higham01,Higham08}, and 
D. Gillespie's classic book~\cite{gillespie91}. 
While this approach is, of course, no substitute for a rigorous 
course on the theory of stochastic processes, it allowed us to present the main ideas of the 
subject succinctly and in a way that was understandable to our interdisciplinary audience.

Following on from this, we focused on applying the core material to research-based topics from across biology. 
These topics were chosen to reflect the interests and expertise of the two instructors and covered models from systems biology, evolutionary game theory, and neuroscience.  
In particular we introduced examples based on biochemical reaction networks, gene regulatory systems, neuronal networks, as well as models of epidemics and evolutionary processes and finally stochastic games and agent-based models. 

This approach allowed us, for example, to motivate continuous Markov processes using the birth-death process, the Gillespie algorithm using
biochemical reaction networks, and point
processes using neuronal spike trains. 

Because of the wide range of topics covered in the course we did not follow a single textbook. However, we offered suggested reading from a number of texts 
covering different topics addressed in the lectures~\cite{allen10,alon19,bower01,bressloff2014stochastic,erban19,gerstner14,nowak2006,wilkinson06}.  
In addition, the computational challenges drew on models published in research papers related to the topic at hand.   
As such, we introduced 
stochastic modeling using a variety of examples from across biology, and provided interested students with a 
groundwork that they could build on in subsequent courses or their own research.



\section{Computational Challenges}
We implemented small-group learning in the course by assigning a sequence of  
computational challenges of increasing difficulty to groups composed of four students: 
Tackling each challenge required the students to work together to understand the background material describing biological and mathematical concepts associated with the problem, and combining both to formulate a 
model. This model then needed to be translated into code contained in a single Jupyter notebook (submitted to a Git repository), with results that could be presented and interpreted in a class discussion. 
As such, the challenges required the students in a group
to share their knowledge and ideas, and learn from each other~\cite{springer99}. 
Once an assignment was completed, each group's code was made available to the whole class. 

Group presentations took place during class, with two groups presenting for 15 minutes each.
Each group gave an overview of the challenge, explained how they approached the problem, 
went over parts of the code, and discussed the results of the simulations. 
Guidelines for presentations were as follows:

\vspace{0.2cm}

\begin{itemize}
\item A brief (1-2 minutes) describing the problem, 
\item 5 minutes explaining how 
they approached and implemented the solution,
\item 5 minutes on results and interpretation
\item A few minutes for question
\end{itemize}

\vspace{0.2cm}

Each group received a grade based on 
the quality of their code, their solution and analysis of the results, and the presentation. The 15 minutes reserved for each  presentation were too
short to give substantive feedback. We therefore held optional, weekly 60 minute Q\&A session to discuss the week's material, including the challenges. 

At the end of the course students gave a short (15 minute) presentation either on a paper related to the subjects discussed in the course, or about their own research, in the style of a short conference talk. If choosing to present a paper, students were asked to verify the results. These final presentations were intended to provide the opportunity for students to practice delivering professional presentations. 

\subsection{Choice of programming language} We required students to implement their models in Python, and use Jupyter notebooks for their presentations. Python is widely used by many researchers as well as in industry, it is flexible and has appropriate libraries for building stochastic models, and it is also free to download and use. In addition, example code for standard procedures is widely available, and we encouraged students to make use of such resources as this is an important skill when writing code for research and other real-word settings. Similarly, by asking students to use a Git repository for their code, we were able to introduce them to a widely used tool for version control and code sharing.

We provided the following instructions to students about coding:
\vspace{2mm}
 \begin{itemize}
    \item You do not have to develop all the code yourself, but you need to understand it. If we don’t say explicitly to implement an algorithm, you can use a package or code that others have written. For example, there are plenty of implementations of the Gillespie algorithm out there. However, make sure you understand what they do. Sometimes the safest thing to do is implementing it yourself.
    \item Find a way to communicate and share code between yourselves. Git is the best for version control, but you may prefer Dropbox. We leave this up to you. You will need to use Git to share the code with us.
    \item Use Slack. We are here to help you. If you need help understanding something, or need a pointer send us a message. You will likely get a reply on the same day, often within an hour, if not from us, then from other students in the course.
\end{itemize}
\vspace{2mm}

Our choice of programming language did create some problems: For instance, some students were not familiar with Python. 
We think that this can be mitigated by communicating these expectations to students well before the
start of the course, as explained above. We therefore believe that Python is a good choice for most modeling courses in applied mathematics.

\section{Example Computational Challenges}  
Here we provide a sample of the computational challenges we assigned along with solutions. Further details about these assignments, model details, and parameters used in the simulation, as well as commented code
showing the solutions, can be found in the accompanying repository. 

We note that as the complexity of the mathematical ideas covered in the lectures increased, the associated challenges became
more open-ended. In particular this meant that there were multiple valid modeling approaches to the same challenge, 
and that there was no  single ``correct'' solution that could be provided in an answer sheet. 
As such, the class presentations and discussions were the primary form of assessment. 

This gradual transition to open-ended problems with more than one acceptable solution was an important part of the course. Beginning graduate students in applied mathematics often find it challenging to accept that in many research problems there may be no single right answer. This problem is compounded by the fact that in most applied mathematics classes, the problems presented do have a single right answer.
The computational challenges allowed us to emphasize the importance of making good modeling choices, looking at limiting cases, and 
asking whether the results of a model make biological sense. 

A good challenge problem should

\vspace{0.2cm}

\begin{itemize}
\item clearly illustrate a mathematical
concept discussed in class, in the context a concrete biological problem;
\item have a solution that offers insight into the behavior of a real biological system;
\item have a solution that can be arrived at and implemented within 10 hours by an average programmer; and
\item have identifiable limiting and test cases where the system's behavior can be predicted either intuitively, or using some straightforward analysis.
\end{itemize}

\vspace{0.2cm}

The three challenges presented below reflect these criteria and illustrate the progression from simple questions with a single concrete solution towards more open-ended complex problems.


\subsection{Rejection Sampling (Challenge 1)}
Generating random numbers and samples from an arbitrary distribution is fundamental to stochastic modeling.
We began the course with a discussion of the numerical techniques for quasi-random number generation and their limitations. We used linear congruential generators (LCGs) to illustrate the ways in which random number generators can fail \cite{Numerical_recipes} 
and introduced the Mersenne Twister as a tool appropriate for the class.

\begin{figure}[!htpb]
    \centering
    \includegraphics[scale=1]{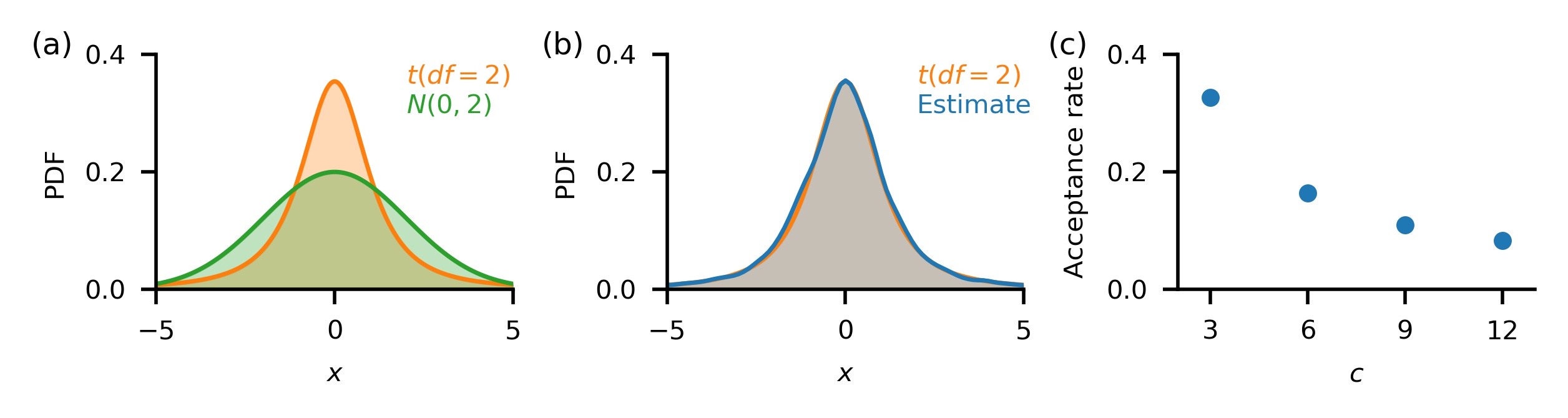}\\
    \includegraphics[scale=1]{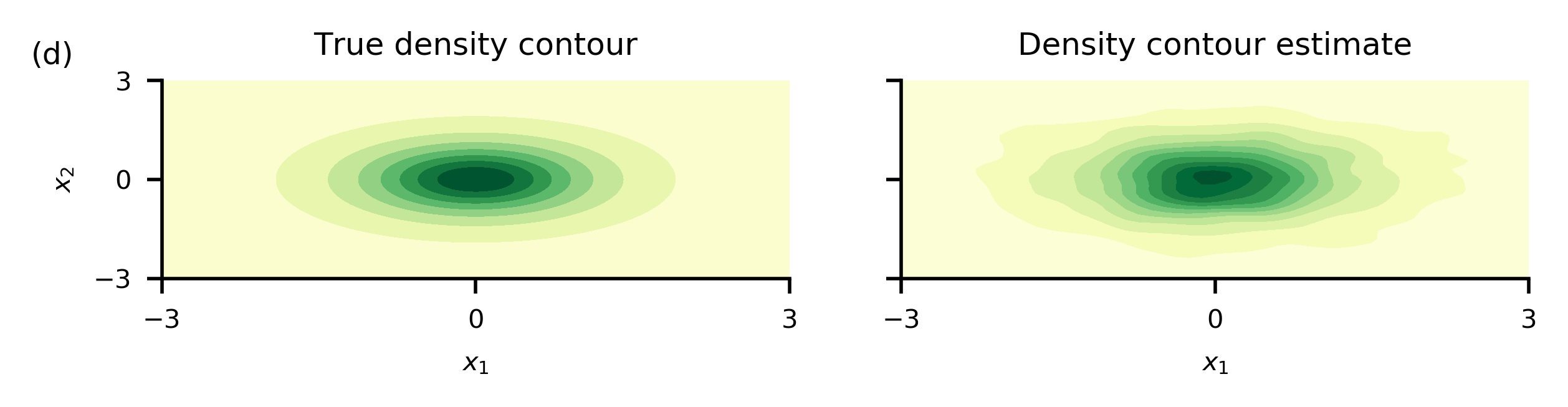}
    \caption{In this example of rejection sampling, the normal distribution was used to generate samples from the $t$-distribution. (a) A comparison of the probability densities of the proposal and target distributions. (b) Kernel density estimate (blue) obtained using 10,000 samples closely matched the target density (orange). (c) Acceptance rates are inversely proportional to the scaling constant, $c$, in the algorithm. (d) In an extension to two dimensions students obtained samples from the bivariate $t$-distribution with parameters $(\mu=[0,0],\Sigma=I_2,df=2)$, using samples from a normal distribution with parameters $(\mu=[0,0],\Sigma=2I_2)$. The contour map obtained using the generated samples (right), shows that the target distribution is well approximated (left).}
    \label{rejection_sampling}
\end{figure}

Our first computational challenge was to use such random number generators to implement rejection sampling and illustrate how random samples from a distribution $f(x)$ can be used to generate samples from a different distribution, $g(x)$~\cite{computing} (See Fig.~\ref{rejection_sampling}). We chose four different combinations of univariate distributions for $f(x)$ and $g(x),$ to produce four different group challenges. We asked the students to visualize the results, and make a convincing case that the samples are generated from the correct distribution. We posed the following questions for discussion: When does the rejection sampling method become inefficient? Do you think that rejection sampling can be extended to higher dimensions? How?

Students solved this challenge easily, and dug deeper: For instance, some looked at how the probability of acceptance of a proposed sample depends on the scaling parameter in the algorithm (See Fig.~\ref{rejection_sampling}c). This scaling parameter serves to ensure that the candidate density has heavier tails than the target, and is optimal at the supremum of the densities ratio. Others implemented the algorithm in higher dimensions (See Fig.~\ref{rejection_sampling}d), allowing us to discuss in more detail how the efficiency of the algorithm depends on the shape of the proposed and target distributions.

While this challenge is simple, it introduced the students to the small-group learning aspect of the class -- to work together to solve the problem and plan a presentation of their results. The simplicity of the problem and its solution allowed us to concentrate on presentation structure and communicate expectations. As the format of the course is different from most others students have attended, the first discussion concentrated on clearly defining what constitutes an acceptable solution to the challenges, and how the solutions should be communicated to the class.

\subsection{The Genetic  Switch and the Repressilator (Challenge 4)}
\begin{figure}[!htpb]
    \centering
    \includegraphics[scale=1]{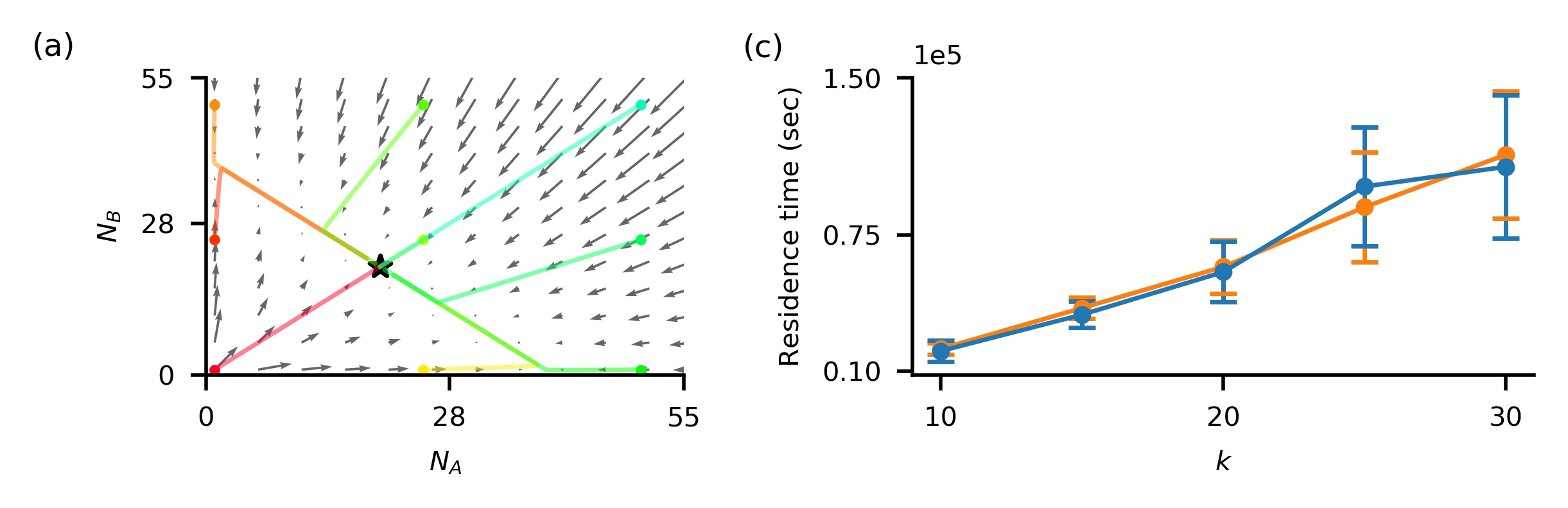}\\
    \includegraphics[scale=1]{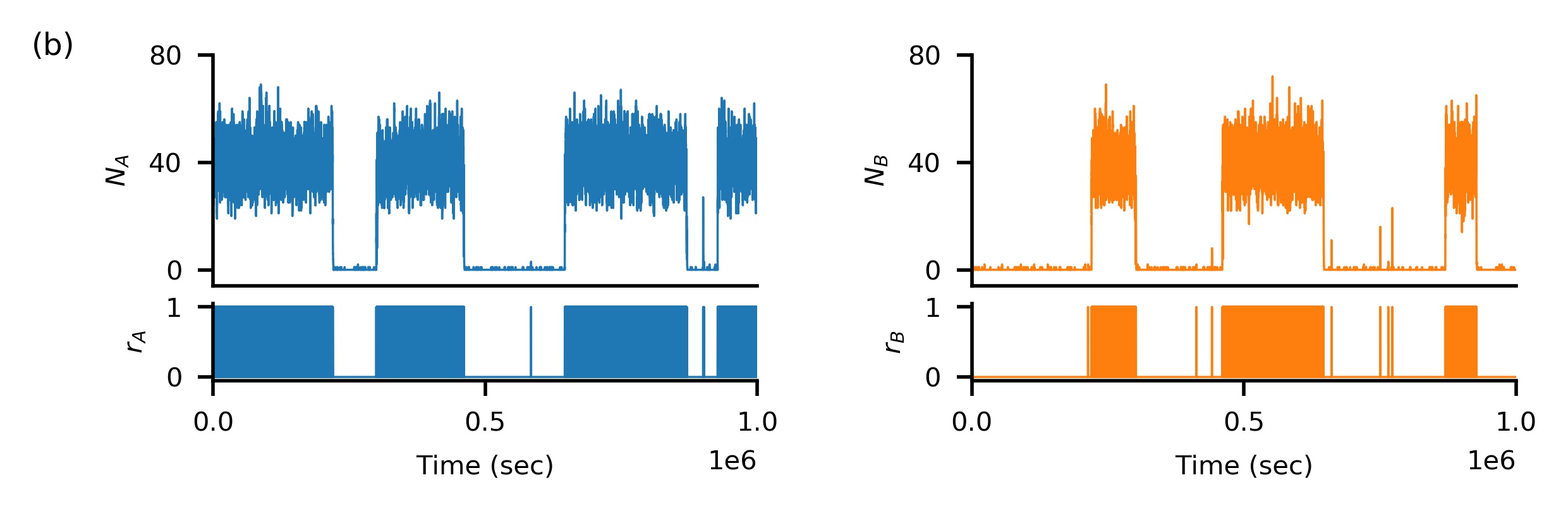}
    \caption{The stochastic exclusive genetic switch of Loinger et al. \cite{loinger07repress} exhibited bistability while its deterministic analog was monostable. (a) A phase plane analysis of the continuous exclusive switch shows a single equilibrium state (black star marker), with a large basin of attraction (circles represent initial conditions). (b) The stochastic model displayed stochastic switching between two complementary states in which one, but not the other, protein was highly expressed. Whether the promoter is bound by either $A$ or $B$, $r_A=1$ or $r_B=1$ respectively, is also indicative of the dominant species as both proteins negatively regulate each other's synthesis. (c) The average residence time in the two states increases with repression strength $k=a_0/a_1$.}
    \label{exclusive_switch}
\end{figure}

The construction of the first synthetic genetic toggle switch~\cite{gardner00} and genetic oscillator~\cite{elowitz00} were landmark events in 
modern biology. The two original papers are lucidly written, and explain how the design of both circuits was inspired by mathematical models. We therefore assigned these two papers as background for this challenge, with optional references for those who wanted to learn more~\cite{cahn14}. 

Although the models presented in these papers are deterministic, genetic circuits often operate at small molecular numbers and are thus inherently noisy~\cite{eldar10}.  
The aim of this challenge was to illustrate how to model noisy genetic circuits, and show how molecular noise can shape their dynamics by inducing state transitions, and oscillations. Stochastic versions of the genetic toggle switch and oscillator are described in a pair of papers by A. Loinger, et al.~\cite{loinger07repress, loinger07toggle}.

Two groups were assigned the genetic switch model, while the other two worked on the represillator. To illustrate the relationship between the deterministic and stochastic description of the system, one group in each pair implemented the deterministic model of the system, while the second group implemented its stochastic counterpart. To simulate the model numerically the second group used the Gillespie algorithm~\cite{Gillespie77} which was the subject of a previous challenge. We therefore also asked the pair of groups assigned to each project to share and discuss their results with one another. 

The first group analyzed the deterministic version of the genetic switch system described in section III in~\cite{loinger07toggle}. They first constructed a Petri net representation of the system as an exclusive switch, and were asked to explain the reason for the term ``exclusive''. They were then asked to solve the ODE numerically and show that for a set of initial conditions the system approached a single equilibrium (See Fig.~\ref{exclusive_switch}a) using phase plane analysis, and computing the equilibrium and determining its stability.

The second group implemented a stochastic model of the same exclusive switch starting with the master equation for the system. They first showed that for the same parameters used in the deterministic model, the stochastic model is bistable, and exhibits noise driven transitions between the two states (See Fig.~\ref{exclusive_switch}b). They next changed a parameter that governs the strength of repression between the two genes in the switch. As repression was increased, so did the average time in each state (See Fig.~\ref{exclusive_switch}c). Both groups were asked to refer to the model equations, and the Petri net to explain these observations.

\begin{figure}[!htpb]
    \centering
    \includegraphics[scale=1]{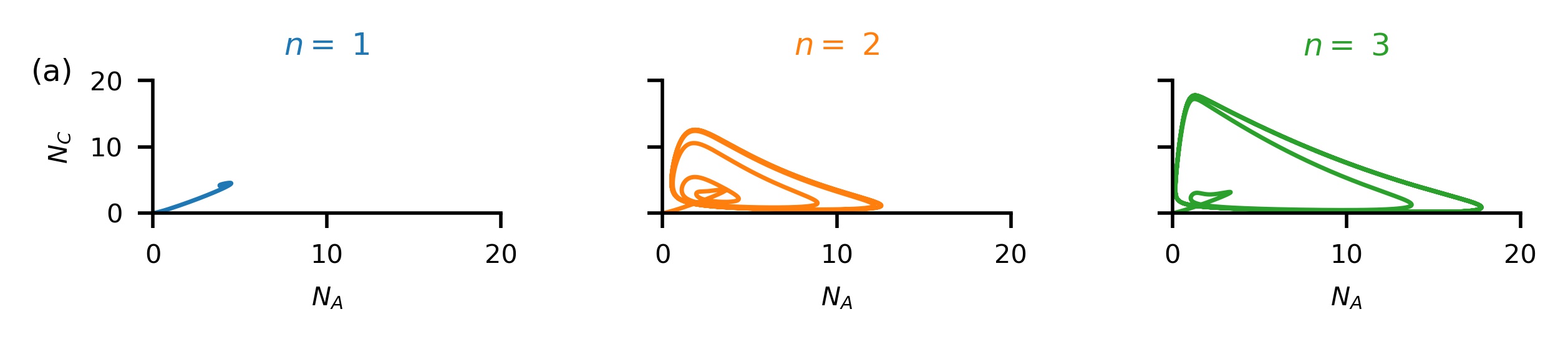}\\
    \includegraphics[scale=1]{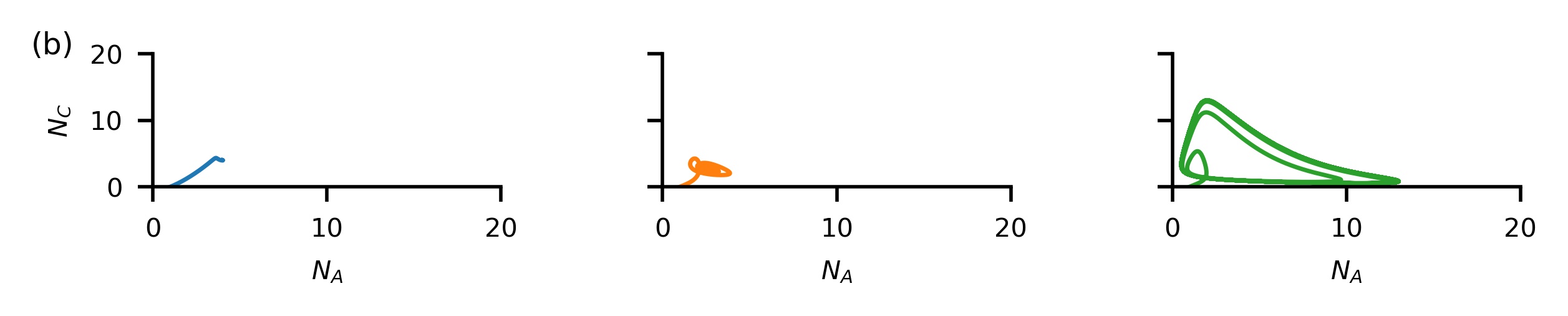}\\
    \includegraphics[scale=1]{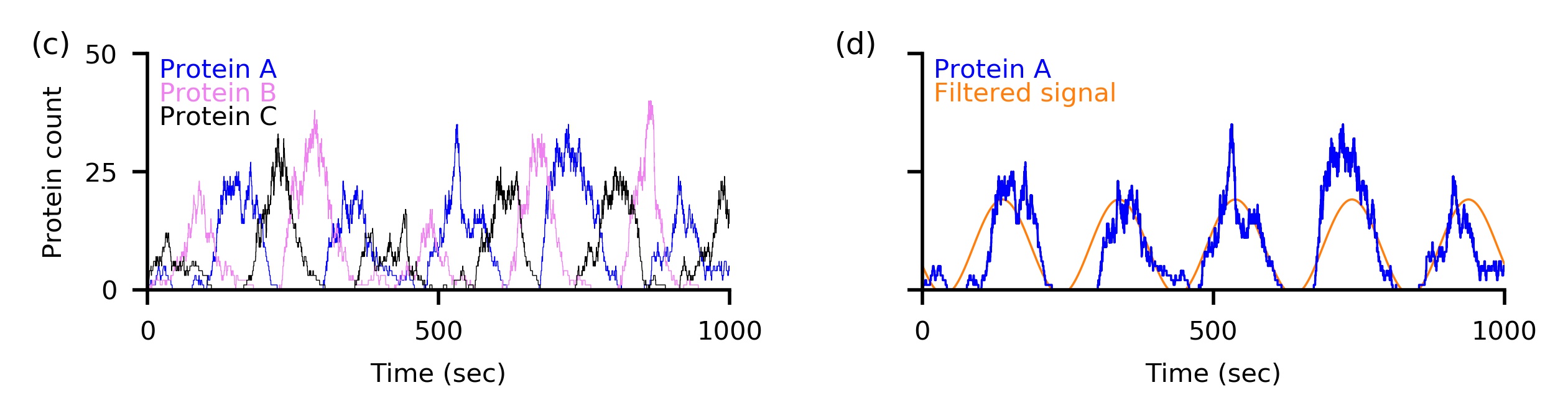}
    \caption{The Michaelis-Menten rate equation models of the repressilator circuit.  The circuit consists of three genes that negatively regulate each other, and can exhibit oscillatory behavior with the three genes expressed in alternation. (a) Simulations show that oscillations appear for Hill coefficient, $n \geq 2$, in the deterministic model that includes mRNA concentrations (Eq. 1 in~\cite{loinger07repress}). (b) However, when mRNA concentration is not modeled explicitly (Eq. 3 in~\cite{loinger07repress}), oscillations occur only when $n=3$. (c) Simulations of the stochastic model including mRNA and using $n=3$, result in oscillations. (d) FFT-based denoising of one of the trajectories, showed a filtered signal allows for an estimate of the oscillation period.}
    \label{represillator}
\end{figure}

The assignment to group three paralleled that of the first group: They implemented and analyze two different versions of a deterministic model of the repressilator~\cite{elowitz00, loinger07repress}. The first version of the model included mRNA dynamics (corresponding to Eq. 1 in~\cite{loinger07repress}),
while the second did not (corresponding to Eq.  in~\cite{loinger07repress}). The inclusion of mRNA level in the model introduced an effective delay in the production of mature proteins, which can be important for generating oscillations. The third group was then asked to show that oscillations in the deterministic system only appear for a high enough value of the Hill coefficient, $n$, which measures the sensitivity of expression (or cooperativity) of each gene to changes in concentration of the regulatory protein (See Fig.~\ref{represillator}a,b).

The fourth group worked in parallel with group three and developed a stochastic model of the repressilator based on Eq. 1 in~\cite{loinger07repress}, and showing that oscillations appear (See Fig.~\ref{represillator}c) consistently for high Hill coefficients, while smaller amplitude, random fluctuations are typical at low Hill coefficient values. We also asked the fourth group to develop a method to detect oscillations and their period by using the FFT of one of the protein concentrations (See Fig.~\ref{represillator}d).


\subsection{Agent-based models (Challenge 9)}
The last challenge of the semester consisted of four group assignments aimed at demonstrating the diverse applications of agent-based simulations in biology. These assignments were the most open-ended of the semester, and required students to apply much of what they learned throughout the course. We asked that they animate their simulations using, for example, \verb|FuncAnimation| from \verb|matplotlib| in Python. For this challenge, each group was given a separate topic based on a classic paper or concept in which space plays a central role in determining the behavior of the system.

{\bf Group 1: Schelling Segregation.} This example of segregation is a classic example of the insight that can be gained by implementing an agent-based model. We provided students with both the original paper introducing the model~\cite{schelling71}, and other more recent discussions of the topic (\emph{e.g.}, see p. 108 in~\cite{easley10}). In the model, two types of agents reside on a lattice, and decide whether to move or stay put according to the number of neighbors of the opposite type in the adjoining spaces. There are several choices that can be made in implementing these rules, but the results are largely independent of those choices. Students were asked to examine the dependence of the final state of the system on the initial ``empty ratio'', the fraction of the domain that is initially unoccupied, and the similarity threshold that determines the fraction of neighbors of opposite type that cause an agent to move. 

Although the Schelling model illustrates a sociological (rather than a biological) phenomenon --  segregation in housing --  similar lattice agent-based models are now used widely in biology~\cite{karamched19,van15}. 

{\bf Group 2: A Spatial Moran Model of Cancer.}  This is an abstract  lattice model of carcinogenesis~\cite{komarova06}. 
Cells are arranged in a regular grid, at locations $i = 0, 1, 2, \ldots , N$. The total number of cells, $N,$ is fixed,
as each cell that dies is replaced by a new cell, with probability determined by the fitness of these cells. Here is a simplified version of the algorithm:
\vspace{2mm}
\begin{enumerate} 
\item[1.] A cell is chosen for death, and is removed from the population. All cells are equally likely to die.
\item[2.] One of the two neighboring cells is chosen for reproduction. If the fitnesses of the two neighboring cells are $r_{\text{left}}$ and $r_{\text{right}}$, the probability that the left will reproduce is $r_{\text{left}}/(r_{\text{left}}+ r_{\text{right}})$, and the probability that the right will reproduce is $r_{\text{right}}/(r_{\text{left}}+ r_{\text{right}})$.
\item[3.] The descendant of the dividing cell fills the empty spot created by the removal of the cell in step 1. 
\end{enumerate} 
\vspace{2mm}

Students were asked to start with different initial positions for a single mutant cell and compute the fixation probabilities (the probability 
that the mutant cell will take over the entire population) for different fitness, and initial position values. They showed the fixation probability against initial position to check for edge effects, and determined the effects of domain size on fixation probability. A master equation for the first version of the model can be found in~\cite{komarova06}.  

{\bf  Group 3: A Model of Leadership and Decision-Making in Animal Groups.} The goal of this challenge was to examine how the decision of informed individuals can impact the behavior of animals in a collective using 
the agent-based model described in~\cite{Couzin05}. In contrast to the previous two models, the domain is continuous and agents can occupy any position in space. The movement of each individual is determined by the motion of its close neighbors. Some of the agents have information about a source of food, or a predator, and adjust their movement by balancing their direction preference and the influence of social interaction. Others are naive, do not have a preferred direction, and only follow their neighbors.

\begin{figure}[!htpb]
    \centering
    \includegraphics[scale=1]{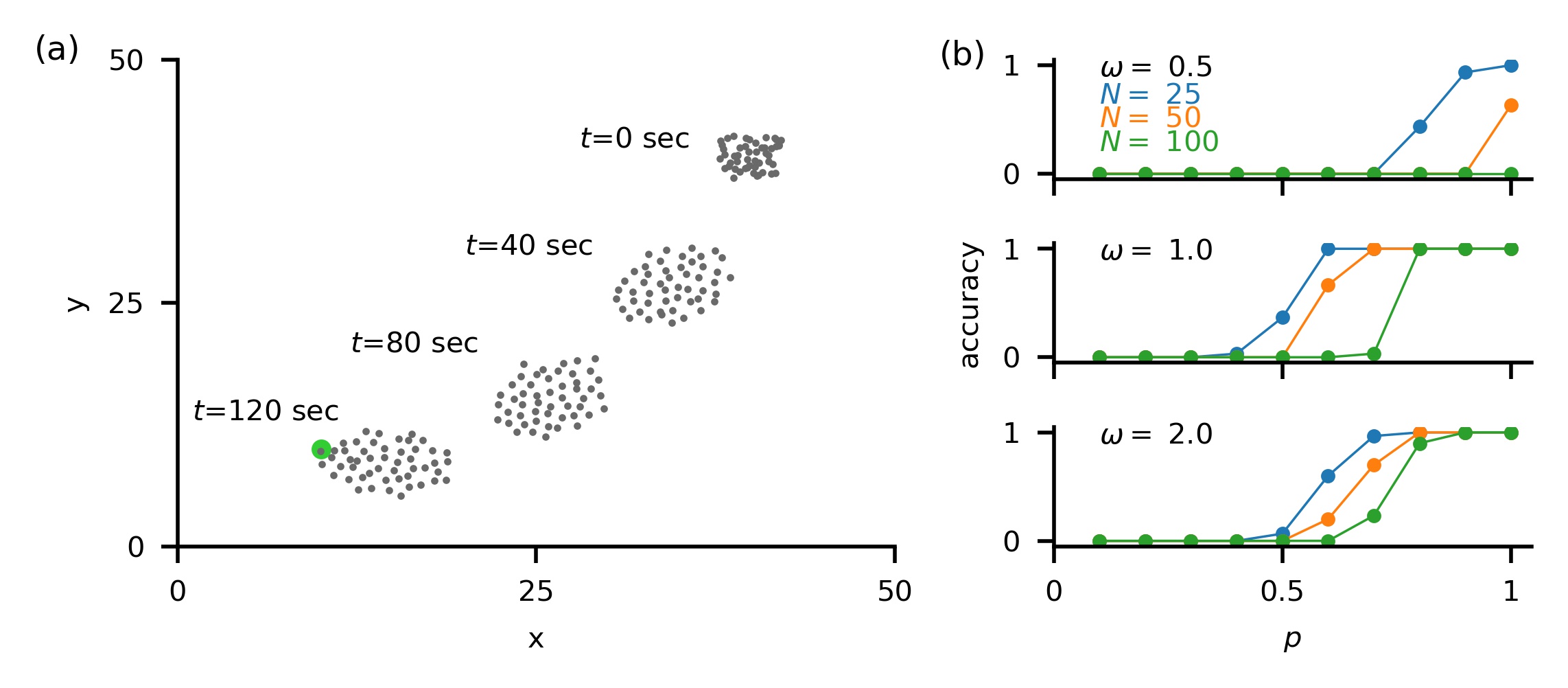}\\
    \includegraphics[scale=1]{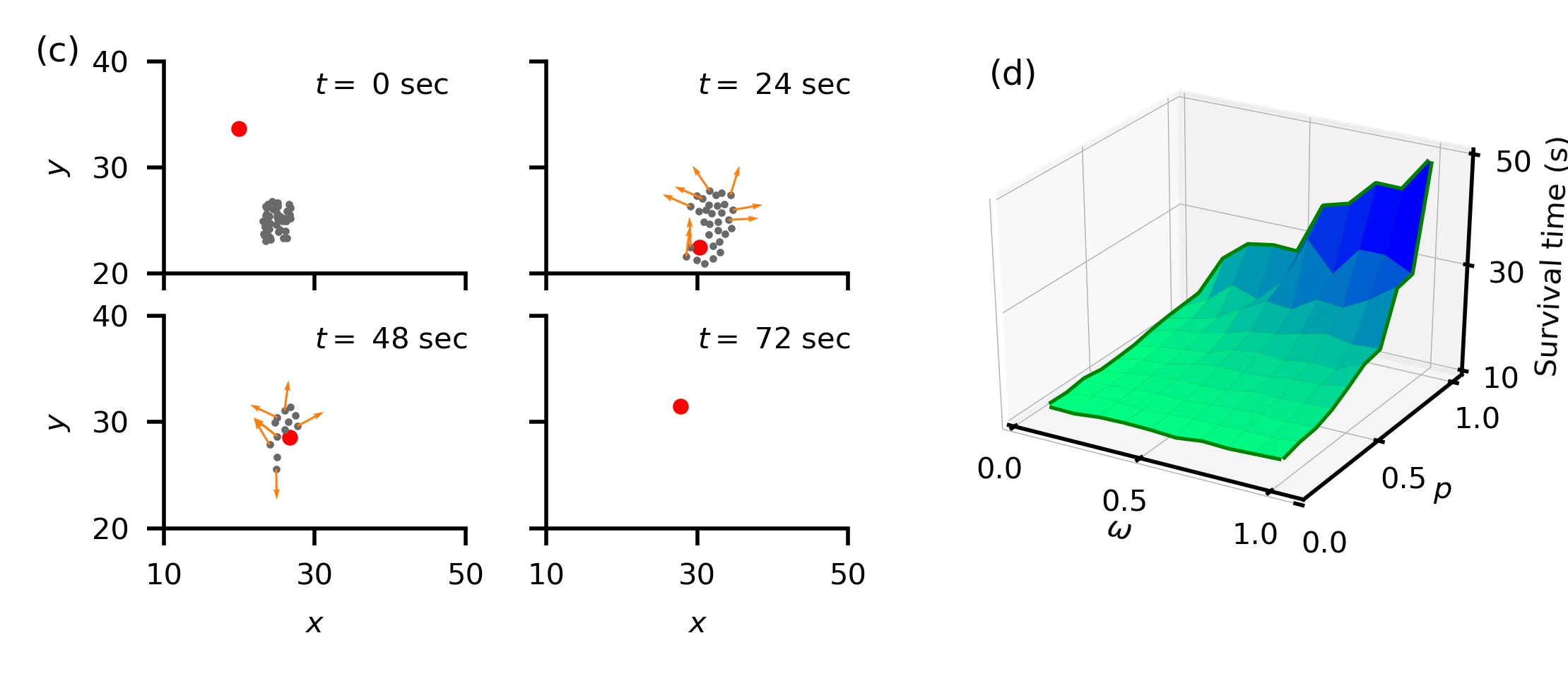}
    \caption{A  model of information transfer in moving groups introduced by Couzin et al. \cite{Couzin05} shows that the strength of social interactions, $\omega$, and the fraction of informed individuals, $p$, are central in a group's success, be it moving to a desired fixed location (a-b), like food and other resources, or escaping from a predator (c-d). Students simulated $N = 50$ agents in a square domain showing that (a) When agents (grey) are informed and social interaction weight is sufficiently high, the group is able to reach a stationary target (green). (b) Group accuracy, quantified by the probability to move to within a 10 units of the target destination within a given time, increases with the fraction of informed individuals, $p$  and social interaction strength, $\omega,$ but decreases with the increase in population size, $N$. (c) Students also simulated escape of the group from a single, fast-moving predator (red). Each agent tries to escape the predator by coordinating with its neighbors, but only some agents have information about the predator's location. Neighboring agents coordinate motion as indicated by their direction (orange arrows). (d) Survival time increased with both interaction strength, and the number of informed individuals.}
    \label{abm_particle}
\end{figure}

In the first part of the challenge, students examined the ability of a collective to find a source of food. They started their simulations with food located at the origin, and placed the group in a small circle far from the food (See Fig.~\ref{abm_particle}a). They then showed that the collective reaches the origin in a time that depends on the number of informed individual.
In the second part of the challenge, students started with populations of different sizes, different numbers of informed individuals, as well as various social interaction strengths. These simulations verified that group accuracy increased with the number of informed agents and the ability of these agents to influence their neighbors, but decreased as population was increased (See Fig.~\ref{abm_particle}b).

The second part of the assignment departed from the original model, as the individuals in the population were attempting to escape a predator rather than find food (see Fig. \ref{abm_particle}c). The predator started at some distance from the population, and always moved to the closest agent, regardless of distance. If the predator came sufficiently close to an agent, that agent was consumed and removed from the population. The predator's speed was slightly larger than the speed of each agent. Fig.~\ref{abm_particle}d shows the survival time of the population as a function of the fraction of informed individuals and the strength of interactions between the agents in the population.   
 
{\bf  Group 4: A Spatial Rock-Paper-Scissors Game.} In this simulation, agents were placed on a lattice with periodic boundary conditions, and each agent used one of the three eponymous strategies~\cite{Reichenbach07}. Each agent thus effectively belonged to one of three species. Each cell (location) in the lattice was initially either empty or occupied by an agent.

Students were asked to start with populating the lattice with agents of each type, and leave a fraction of cells empty. Simulation proceed by picking a cell uniformly at random in the lattice, and then picking a random neighboring cell. If both cells are empty, or occupied by agents of the same species, nothing happens. Otherwise three things can happen:

\vspace{2mm}
\begin{enumerate}
\item[1.] If only one of the cells is occupied, the agent in the occupied cell reproduces. The descendant belongs to the same species, \emph{i.e.} uses the same strategy.
\item[2.]  If the two cells are occupied by different species, then with probability $p$ they fight. The agent with the loosing strategy dies, and their cell is vacated. 
\item[3.] With probability $1-p$, the agents in the two cells swap places.
\end{enumerate}
\vspace{2mm}

\begin{figure}[!htpb]
    \centering
    \includegraphics[scale=1]{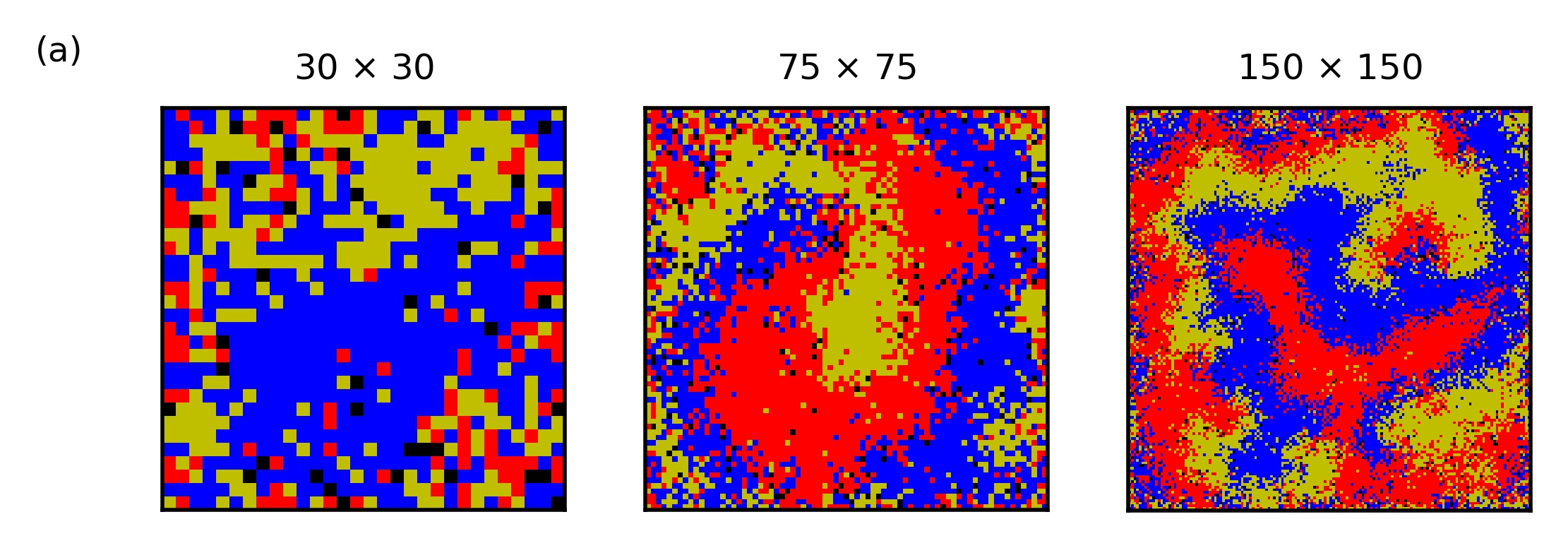}\\
    \includegraphics[scale=1]{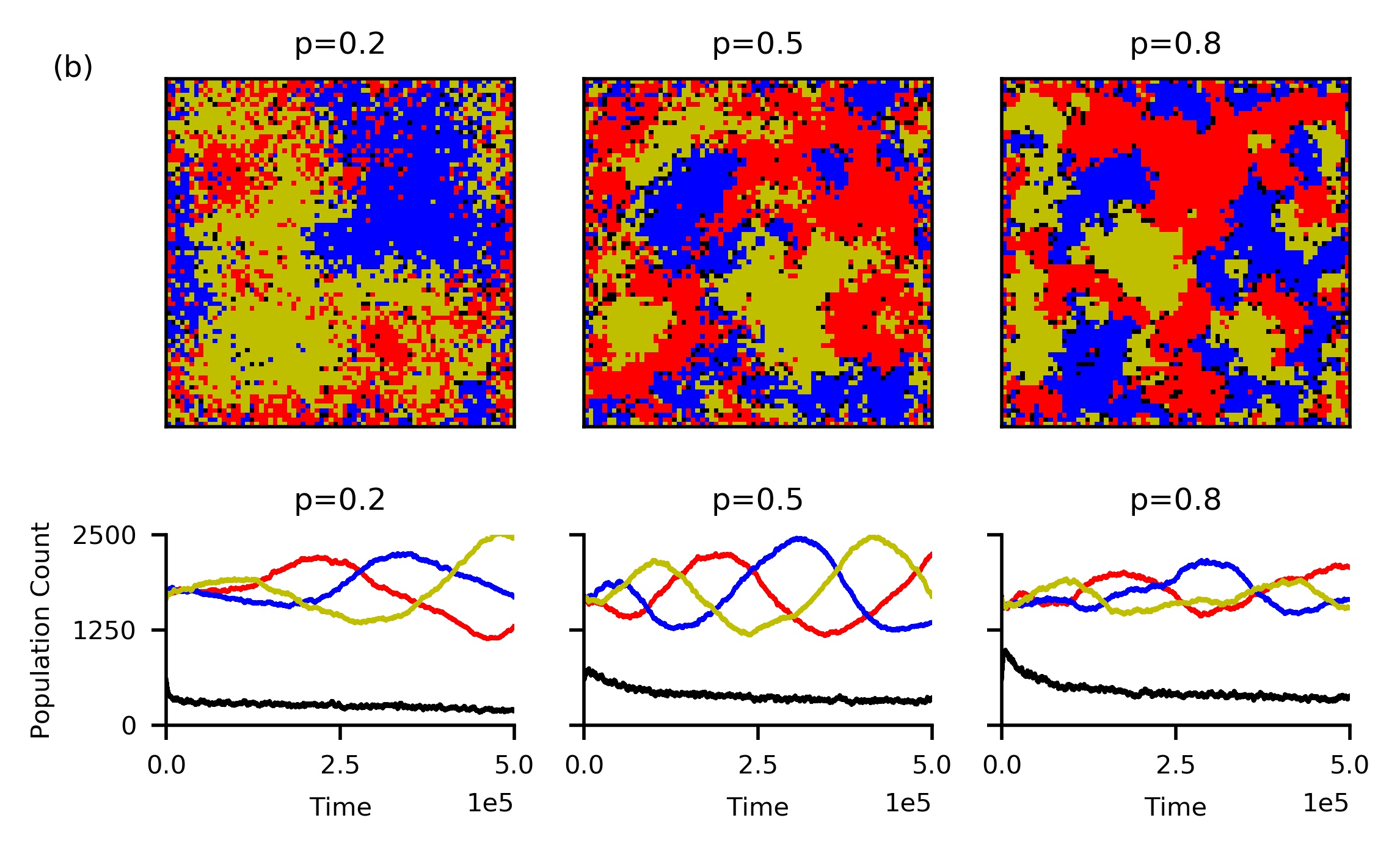}
    \caption{Lattice size and mobility strongly affect population dynamics in a spatial rock--paper--scissors game. In all simulations, a cell in the lattice was initialized to be empty (black) with probability 0.1, and belonged to one of the species (red, blue, yellow) with probability 0.3, respectively. (a) Games on  square lattice of different sizes, $30 \times 30,75 \times 75$ and $150 \times 150$, with mobility probability, $1-p$ for $p=0.2$, showed that a single species is likely to dominate in a small domain. Coexistence is more likely in larger domains where spiral patterns developed, consistent with the observation of Reichenbach et al. \cite{Reichenbach07} (b) In a lattice of size $75 \times 75$ students observed oscillation in population fractions with frequency decreasing, and cluster size increasing with $p$.}
    \label{lattice}
\end{figure}

The three strategies can coexist in a sufficiently large domain (See Fig.~\ref{lattice}a) and the dynamics of the system changes significantly with the parameter $p$ (See Fig.~\ref{lattice}b). Mobility, modeled through agents swapping positions with probability $1-p$, is crucial to population diversity, as species form spiraling spatial patterns that grow in size with mobility (See Fig.~\ref{lattice}a,b).
Over a wide range of parameters, the fraction of each species oscillates through time, at a period that depends on $p$ (See Fig.~\ref{lattice}c). A possible interesting extension of the model, which we left as a bonus challenge, was to simulate five species by using the rules of the rock-paper-scissors-lizard-spock game.

\section{Group Interactions}

The benefits of collaborative learning are well-doc\-umented~\cite{johnson2000cooperative,prince2006inductive}. Problem-based learning is widely used in some sectors such as medical education \cite{Barrows80,Mansur14,Neville09,Smits02}, 
 however it is less common in mathematics and STEM education more generally. Solving problems in 
groups allows students to learn inductively, as well as develop communications skills. The
ability to work as a member of a team is useful regardless of their career path. 
A succinct and extremely 
valuable introduction to this topic is provided by of Richard M. Fleder and collaborators~\cite{felder2001effective,oakley2004turning}, and there are many other excellent books
on the subject~\cite{millis1997cooperative}. Here we focus on the issues we encountered implementing small-group learning in a graduate course on mathematical modeling aimed at an interdisciplinary audience.

For each challenge we randomly assigned 
students to teams of four, and selected them to choose a person who will present 
their work. However to ensure all students participated in presenting, only students with the least number of presentations to date were 
able to present in any given week. The strategy for choosing a presenter worked well allowing 
all students to present about the same number of times during the semester. However, we found that reshuffling groups for each assignment was disruptive and made it harder to assess contributions and ensure equitable division of labor in the long run. In future, we plan to keep group composition constant throughout the semester
with initial group assignments informed by responses to a short questionnaire about skills and background. This will allow us assign teams of students with
heterogeneous skill sets.   
Assigning groups prevents homogeneity in which math students seek out other math 
students, etc. Such homogeneity is not desirable in an interdisciplinary group since it puts some groups at a disadvantage and prevents knowledge exchange 
in addition to the loss of the other benefits of interdisciplinary groups~\cite{oakley2004turning}. 

The biggest challenge we faced with small-group learning was maintaining a fair division of labor in the computational challenges. 
Students were asked to fill out feedback forms after each challenge, with many noting this problem repeatedly. However students also found providing feedback weekly onerous. In future we will collect more detailed feedback after every three challenges.


There are several collective learning strategies that can be implemented to improve the 
success of small-group learning in our course: Providing a \emph{Team Policy Statement} and \emph{Team Expectation Agreement} at the outset of the course will provide a better basis for collaboration 
regardless of the course topic~\cite{oakley2004turning}. In particular, rather than just 
the presenter, assigning roles to all members of the group would make sure that 
everyone is accountable and less able to free ride: e.g. assigning a \emph{coordinator} to organize meetings, and keep the group
on task, a \emph{recorder and presenter} to prepare the final presentation that will be delivered to the class, a \emph{monitor} who checks that everyone agrees with the final results, and a \emph{checker} who checks the final results. Finally, if meeting in person, we would institute a group office hour \emph{before} the assignment is due to go over the
solution and the presentation with the students.


\section{Student Feedback}

At the end of the course we administered an exit survey, which we summarize the results of here. All questions and full results are available at the repository accompanying this article.

The 14 students who responded overwhelmingly agreed that the computational challenges helped them understand the 
course material. They were somewhat more divided about whether what they learned was
directly applicable to their research, although the majority agreed. Most students 
who answered (8/14) indicated that they agreed or strongly agreed with the statement that 
they would take a course with computational challenges again. Most students strongly agreed
with the statement that the course helped improve their coding skills. 

Students were more critical about the division of labor within the groups. There was a weak 
majority who thought that there were too many challenges. This is really something that depends on the composition of the class. We asked them to self report the time they spent on the challenge (3 reported 2--4 hours, 6 reported 4--6 hours, and 7 reported more than 6 hours per week). We note that all groups completed all challenges. Although the quality of the results varied, they generally exceeded our expectations.
A better guide to division of labor, as discussed above, would likely improve this.
Students were largely positive about small-group learning, suggesting that problem-based learning can be successful in the context of an applied mathematics course. 

There was a diversity of responses to a question about the time spent on class discussions of computational challenges. Mandatory in person office hours prior to the time the assignment is due would be useful in addressing this, and is not unduly onerous in graduate courses where enrollment 
is smaller, although it may require the help of a teaching assistant.

\section{Conclusions}

Designing and integrating a computational component into a course on applied mathematics
can be time consuming, however we observed clear benefits to 
both the students and the instructor.
Small-group learning, particularly with an interdisciplinary student cohort, also provided clear benefits, despite the challenges with ensuring equitable division of labor, and we plan to continue using this approach in future. This type of course is especially relevant for graduate students. Science is becoming increasingly collaborative.
Whether our students continue in academia or industry, they will likely have to work in teams, and these teams
are likely to include far greater diversity of skill and knowledge than the groups in this course. For many, their future success will thus depend on how well they 
can work as part of such groups. 

While our course covered many topics, several were only touched upon. While we asked students to 
provide a well motivated analysis of the results, and we provided feedback, we
only spent the last week of the course on fitting models to data~\cite{wilkinson06}. 
A companion course, also including an element of problem-based learning, and focused on model fitting, validation, and data 
driven model discovery and assimilation would be an excellent complement to the one described here. 
Such a course could be taught using publicly available, curated datasets.  
As these are areas of high interest and intense research activity,  
there are now excellent resources that could be used to develop such a course~\cite{brunton2019data}.

Throughout the semester we effectively provided four 25 minute lectures per week that we could easily have pre-recorded. Alternatively, we could have used existing resources, such as the videos~\cite{erban} accompanying the introductory book to stochastic processes in biology~\cite{erban19}.
This would have allowed us to ``flip'' the classroom~\cite{berrett12}, spend less time in class on instruction. Students were most engaged when 
we discussed their modeling 
and implementation choices, and checking and interpreting their results. They were invested in the challenges, often spending considerable
time each week.



A course making use of examples from current research can make effective use of crowdsourcing. We welcome help and suggestions to expand on the set of challenges we have made available. Developing the challenges, 
seeing the different, often innovative ways in which students solved them, and discussing the solutions was very rewarding for the instructors. We therefore hope to spur the development of similar courses across different branches of applied mathematics.

\section*{Appendix A. Repository of resources}
We provide a detailed syllabus, homework, and the text of all computational challenges, along
with code for the examples we discussed here at ~\url{https://github.com/josic/stochastic_process_bio}. This repository will be maintained.

\section*{Acknowledgments}
We would like to thank the students in the course for their enthusiasm and willingness to participate 
in an experimental course, all during a pandemic. This work was supported by the National Science Foundation (NSF) through the grants
DMS-1662305 (KJ,MJC), MCB-1936770 (KJ), DBI-1707400 (KJ,AEA), the National Institutes of Health grant RO1 GM117138 (KJ,MJC).

\end{document}